\documentclass[aps,prb,twocolumn,groupedaddress,showpacs,showkeys]{revtex4}
\usepackage{graphicx}
\usepackage{dcolumn}
\usepackage{bm}
\begin{document}
\preprint{APS/123-QED}
\title{Excitations of the $\nu=5/2$ Fractional Quantum Hall State and the Generalized Composite Fermion Picture}
\author{George E. Simion}
\author{John J. Quinn}
\affiliation{Department of Physics and Astronomy, University of
Tennessee}

\date{\today}

\begin{abstract}
We present a generalization of the composite Fermion picture for a
muticomponent quantum Hall plasma which contains particle with
different effective charges. The model predicts very well the
low-lying states of a $\nu=5/2$ quantum Hall state found in
numerical diagonalization.

\end{abstract}

\maketitle

%%%%%%%%%%%%%%%%%%%%%%%%%%%%%%%%%%%%%%%%%%%%%%%%%%%
%%%%%%%%%%%%%%%%%%%%%%%%%%%%%%%%%%%%%%%%%%%%%%%%%%%
\section{Introduction}

The energy spectrum of a pure two-dimensional electron gas in a
strong perpendicular magnetic field is completely determined by the
Coulomb interactions, making it the paradigm for all ``strongly
interacting" systems, (for which standard many body perturbation
theory is inapplicable). Jain's composite Fermion (CF) picture
\cite{JainPRL89, JainPRB90} has emerged as a comprehensive approach
to understanding the most prominent incompressible states of the
lowest Landau level (LL0). The original CF picture made use of a
mean-field Chern-Simons (CS) gauge field theory. The CFs are
electrons that have an even number Chern-Simons (CS) fluxes
attached. The remarkable feature of this model is that it maps
complicated FQH states with filling factor $\nu$ into simpler and
better understood states of CFs with effective filling factor
$\nu^{\ast}$. A good example is provided the well-known series of
Laughlin-Jain states \cite{LaughlinPRL83, JainPRB89} occurring at
filling factors $\nu=\nu^{\ast}/(2p\nu \pm 1)$ where $\nu^{\ast}$
and $p$ are integers. The mean-field approach deals with two  scales
of energy: the cyclotron energy $\hbar \omega_c $ and the Coulomb
interaction scale $e^2/\lambda$ where the cyclotron frequency is
$\omega_c=(eB)/(mc)$ and the magnetic length is $\lambda
=\sqrt{(\hbar c)/(eB)}$. In numerical studies and in some realistic
systems, the cyclotron energy is much larger than the average
Coulomb energy; thus the former scale is irrelevant. The low lying
energy states are completely determined by Coulomb interactions.
Adiabatic addition of flux \cite{QuinnQuinnPRB03} introduces
Laughlin correlations without the need of a new mean-field cyclotron
energy scale. Generalization of the CF picture to a plasma
containing more than one type of particle was first introduced by
W\'{o}js et al. \cite{WojsSzYiQuinnPRB99}. It was used to predict
the low lying band of states of systems containing both electrons
and valence band holes. In this case the the constituents (electrons
and negatively charged excitons) have the same charge. In this paper
we propose a new generalization of CF model to include
multi-component plasmas in which particles with different charges
are involved. We use this model to suggest an explanation of the
low-lying bands of states of the $\nu=5/2$ FQH states.

The $\nu=5/2$ state has generated considerable recent interest as
the result of the suggestion that its elementary excitations can be
non-Abelian quasiparticles. Such excitations occur when the three
particle pseudopotential $V(\mathcal R_3)$ of Greiter et
al.\cite{GreiterWenWilczekNuclPhys92}, which forbids the formation
of compact three particle clusters, is used to describe the
interactions of electrons in the first excited Landau level (LL1).
Our simple generalized CF picture is used to interpret the ground
state and low lying bands of excitations of electrons in LL1
interacting through the standard Coulomb pseudopotential appropriate
for such electrons confined to a very narrow quantum well. New
experimental results of Choi at al. \cite{choiPRB08} have been used
to interpreted incompressible states in terms of formation of pairs
(with $\ell_{\rm{P}}=2\ell-1$) or large clusters at filling factors
$\nu_1=\nu-2$ of LL1 with $2/3 \geq \nu_1 \geq 1/3$. The most
prominent IQL state occurs at $2\ell=2N-3$ (or at $2\ell=2N+1$).
Here we present numerical results for the energy spectrum for
$N=8,10,12$ and $14$. We demonstrate that in addition to the $L=0$
ground states, the two lowest band of excitations can be interpreted
using our simple generalization of Jain's CF picture to a two
component plasma of Fermion pairs (FPs) and unpaired electrons.

The paper is organized as follows. The next section is dedicated to
the solution of a two particle system in a magnetic field. The third
section discusses CS gauge transformation and adiabatic addition of
CS flux. In the fourth section, we present the generalized CF model
and its application to the $\nu=5/2$ FQH state. We draw conclusions
in the last section.

%%%%%%%%%%%%%%%%%%%%%%%%%%%%%%%%%%%%%%%%%%%%%%%%%%%
%%%%%%%%%%%%%%%%%%%%%%%%%%%%%%%%%%%%%%%%%%%%%%%%%%%%

%%%%%%%%%%%%%%%%%%%%%%%%%%%%%%%%%%%%%%%%%%%%%%%%%%%%%%
%%%%%%%%%%%%%%%%%%%%%%%%%%%%%%%%%%%%%%%%%%%%%%%%%%%%%%
\section{Two charged particles in a magnetic field}

A two particle system is the simplest system that can be used to
understand the physics of introducing the CS flux. The two particles
have masses $m_1$ and $m_2$ and charges $q_1$ and $q_2$ and they are
moving in the $x-y$ plane. A dc uniform magnetic field $\vec B=B\hat
z$ is applied perpendicular to the plane. We work in the symmetric
gauge where the vector potential is $\vec A =  \frac{1}{2}Br \hat
\phi$, where $\hat \phi$ is the unit vector in the direction of
increasing the angular coordinate $\phi$ . The Hamiltonian contains
the kinetic terms and the Coulomb term.
\begin{widetext}
\begin{equation}
\label{eq:genH} \hat H=\frac{1}{2m_1}\left[\vec p_1-\frac{q_1}{c}
\vec A (\vec r_1)\right]^2+\frac{1}{2m_2}\left[\vec p_2
-\frac{q_2}{c} \vec A (\vec r_2)\right]^2+\frac{q_1 q_2}{|\vec
r_1-\vec r_2|}~.
\end{equation}
\end{widetext}

The Hamiltonian can be separated into the center of mass (CM),
relative (R) part, and a part describing the interaction between the
CM and R motions (I).
\begin{widetext}
\begin{eqnarray}
\label{eq:HCM} H_{CM}&=&\frac{1}{2M}\left[\vec P
-\frac{Q}{c}\vec A(\vec R)\right]^2+ \frac{\tilde q^2 \vec A^2(\vec R)}{2\mu c^2}~, \\
\label{eq:Hrel} H_R&=&\frac{1}{2\mu}\left[\vec p
-\frac{q}{c} \vec A (\vec r)\right]^2+\frac{q_1q_2}{r}+\frac{\tilde q^2 \vec A^2(\vec r) }{2Mc^2}~,\\
\label{eq:Hint} H_I&=&-\frac{\tilde q}{\mu c}\left[\vec A(\vec R)
\vec p +\vec P \vec A (\vec r)\right]+\frac{\tilde q \bar q}{2 \mu}
\vec A (\vec r)\vec A(\vec R)~.
\end{eqnarray}
\end{widetext}
Here relative coordinate and momentum are $\vec r=\vec r_1 - \vec
r_2$ and $\vec p=\frac{1}{2}(\vec p_1-\vec p_2)$, respectively. The
reduced mass and charge are  $\mu=(m_1 m_2)/(m_1+m_2)$ and
$q=(q_1m_2^2+q_2m_1^2)/(m_1+m_2)^2$. $R=\frac{1}{2} (\vec r_1+\vec
r_2)$ and $\vec P=\vec p_1+\vec p_2$ represent the CM coordinate and
momentum. Also $\bar q=(q_1m_2+q_1m_2)/(m_1+m_2)$ and $\tilde
q=(q_1m_2-q_2m_1)/(m_1+m_2) $. Throughout this paper we consider
that the particles have the same charge to mass ratio (specific
charge) $q_1/m_1=q_2/m_2$, and as a consequence the $\tilde q=0$ and
the CM and R motions decouple.

\begin{widetext} The solution of Schr\"{o}dinger equation of the relative
motion can be written as $\Psi_{nm}=e^{im\phi}u_{nm}(r)$. Exchanging
the particles corresponds to replacing $\phi$ by $\phi+\pi$.
Therefore, if particles are Fermions (Bosons), $m$ must be odd
(even). The radial function $u_{nm}(r)$ satisfies the following
equation
\begin{equation}
\left[-\frac{\partial ^2}{\partial \rho^2}-
\frac{1}{\rho}\frac{\partial}{\partial \rho}+ m +\frac{m^2}{\rho^2}
+\frac{\rho^2 }{4}+  2 \frac{q_1q_2}{q^2}\frac{\eta_0}{\rho} \right]
u_{nm}(\rho)=\epsilon_{nm} u_{nm}(\rho)~,
\end{equation}
where $\rho=r/ \lambda_r$, $\lambda_r=[(\hbar c)/(qB)]^{1/2}$ is the
magnetic length for the relative motion. The parameter $\eta_0$
characterizes the strength of Coulomb interaction with respect to
cyclotron energy. $\eta_0=(q^2/\lambda_r)/(\hbar \omega_{cr})$ and
$\omega_{cr}=(qB)/(m c)$ is the cyclotron frequency of the relative
motion. The eigenvalues of $H_R$ are $E_{mn}=(\hbar \omega_{cr}/2)
\epsilon_{nm}$. $n=0,1,2,\cdots$ is a non-negative integer number,
while $m$ is an integer less than or equal to $n$.
\end{widetext}

In the absence of the electron-electron term the solutions is
well-known
\begin{equation}
\label{eq:wf_nint}
u_{nm}^0(\rho)=\sqrt{\frac{n!}{2^m(n+|m|)!}}\rho^{|m|}L_n^{|m|}(\rho^2/2)\exp(-\rho^2/4)~.
\end{equation}
The lowest Landau level wavefunction has its maximum value at
$\rho=\sqrt{2|m|}$. The eigenvalues are $\epsilon_{nm}=2n+1+m+|m|$.

\begin{figure}
\centerline{\includegraphics[width= 0.5 \linewidth]{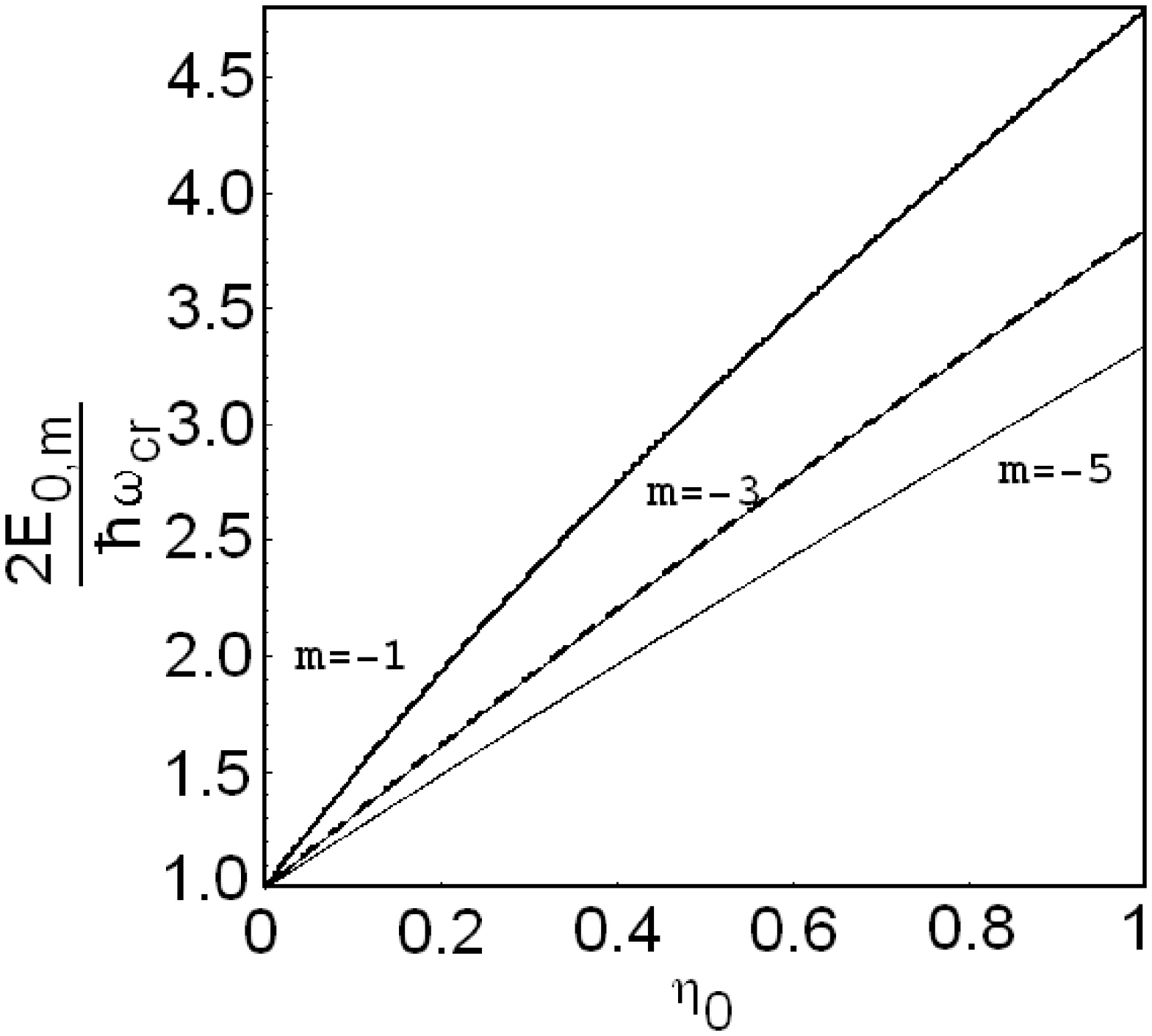}
\includegraphics[width= 0.5 \linewidth]{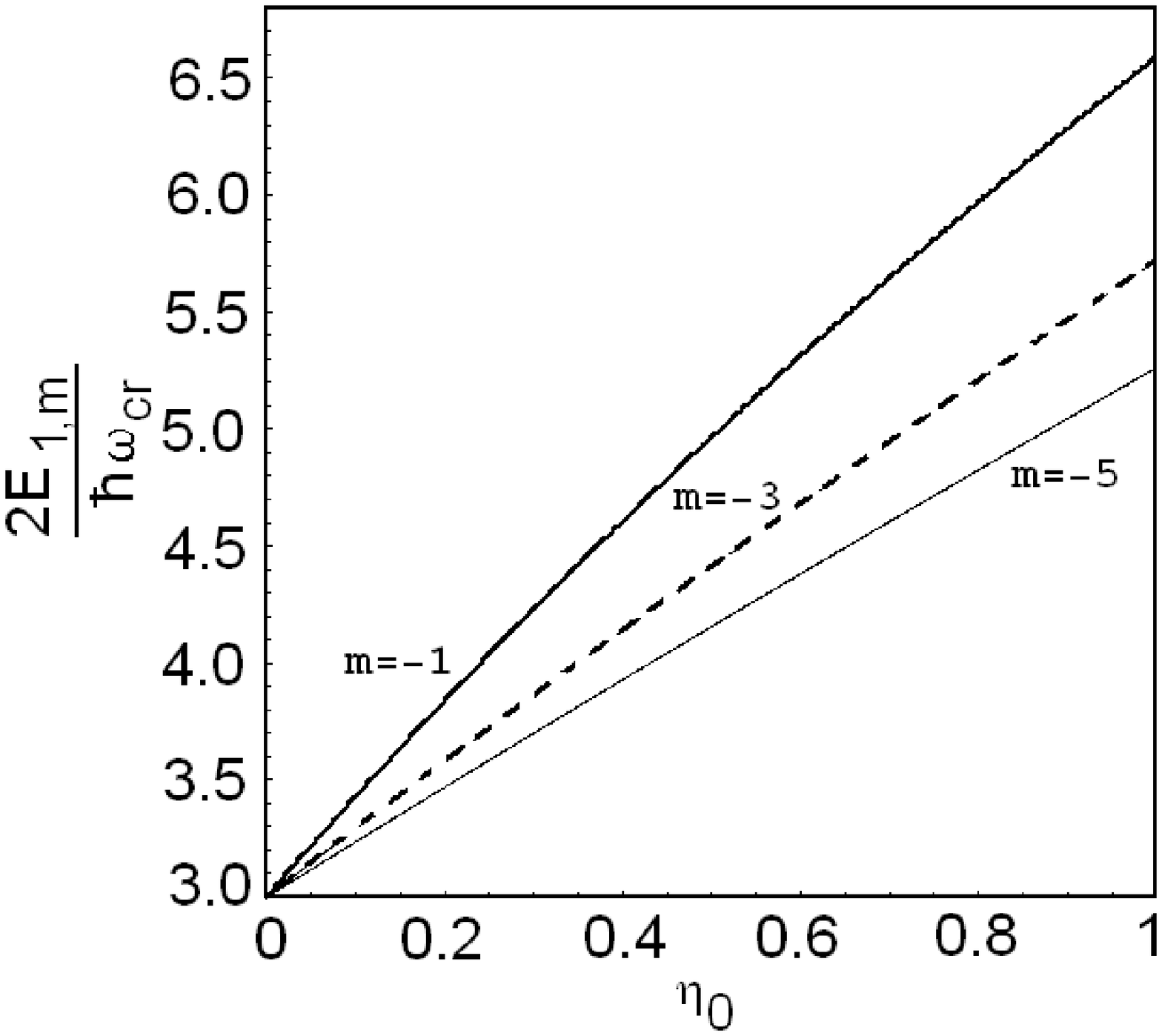}}
\caption {Energy of the relative motion of the two particle system
in LL0 (left panel) and LL1 (right panel) as a function of the
interaction strength parameter $\eta_0$.} \label{Fig:Enmint}
\end{figure}

\begin{figure}
\centerline{\includegraphics[width= 0.5
\linewidth]{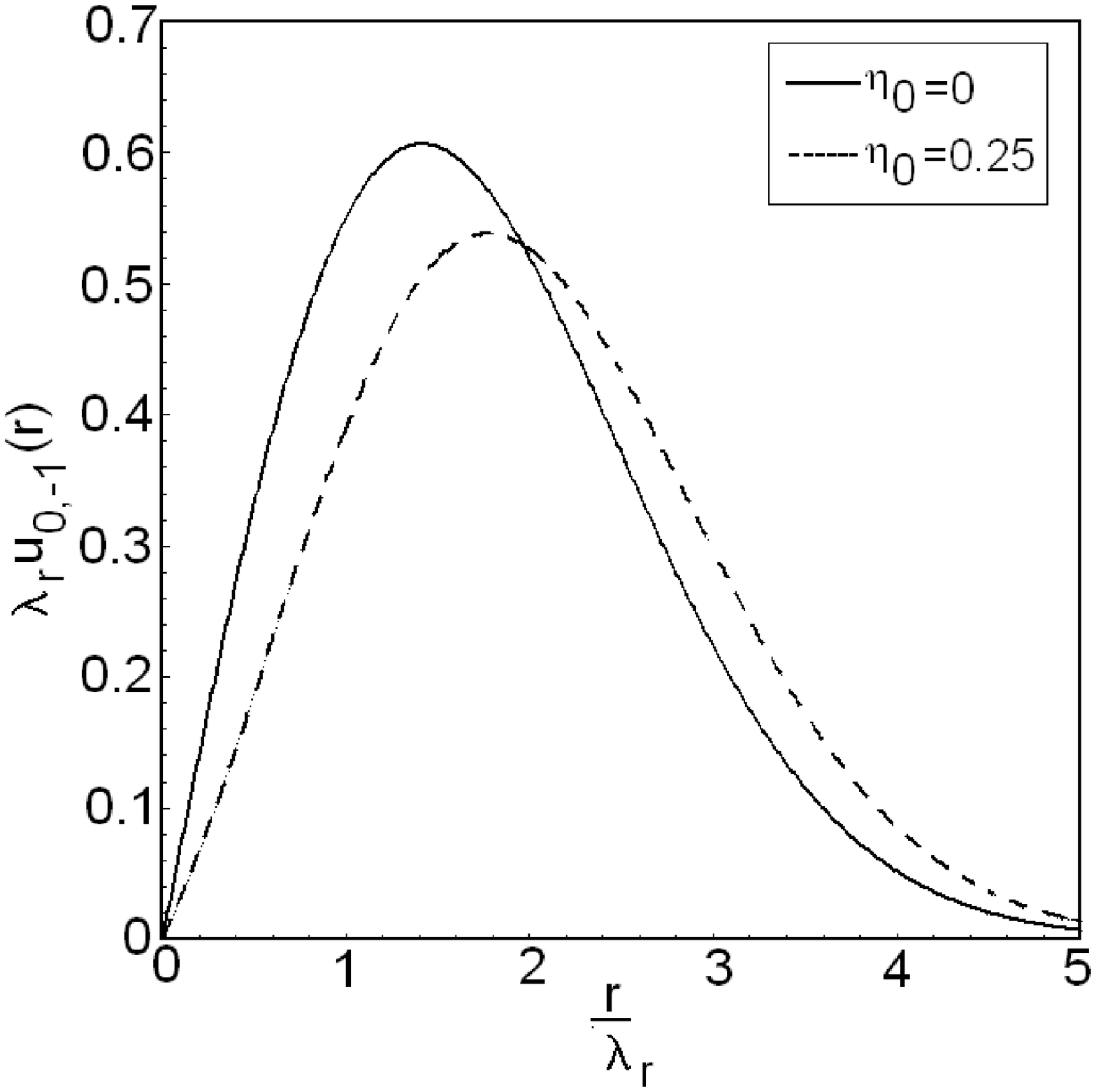} \includegraphics[width= 0.5
\linewidth]{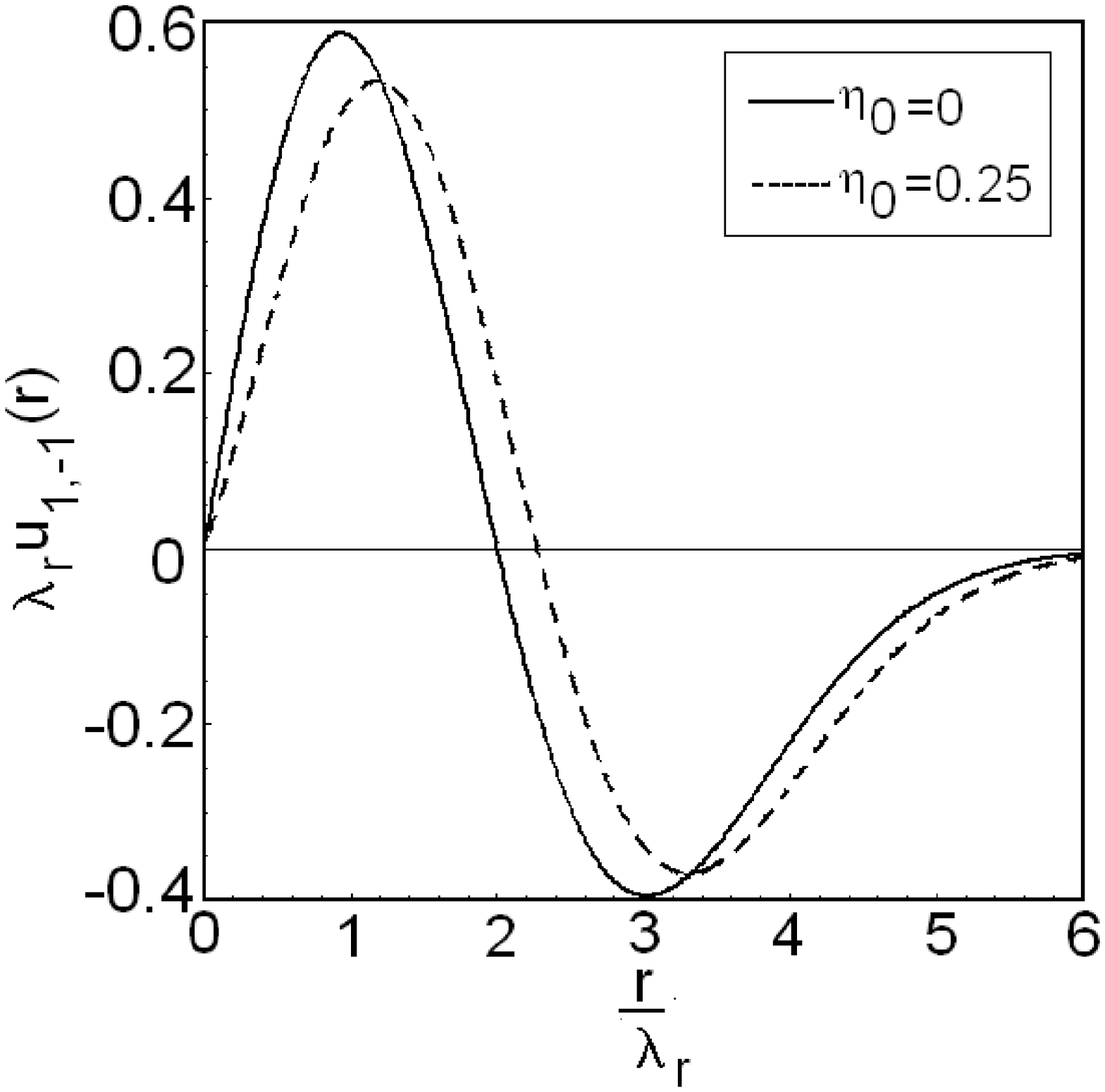}} \caption{Radial part the relative
motion wavefunction of the two particle system in LL0 (left panel)
and LL1 (right panel) for $\eta_0=0$ (independent particles) and
$\eta_0=0.25$.} \label{Fig:Wavefunction}
\end{figure}

If the Coulomb interaction is taken into account, the degeneracy of
the Landau levels is lifted. States with small values of $m$ have
the largest increase in energy. In Fig. \ref{Fig:Enmint}, we plot
the energy levels vs. the relative strength of Coulomb interaction
for a system of 2 electrons. Coulomb interaction does not alter
fundamentally the shape of the wavefunction at least for small
values of $(e^2/\lambda_r)/(\hbar \omega_{cr})$, preserving the
small and large distance behavior. The radial part of the
wavefunction for $\eta_r=0.25$ together with the corresponding one
for noninteracting particles are plotted in Fig.
\ref{Fig:Wavefunction} for the same two-electron system.

%%%%%%%%%%%%%%%%%%%%%%%%%%%%%%%%%%%%%%%%%%%%%%%%%%%%%%%%%%%%%%%%%%%%%%%
%%%%%%%%%%%%%%%%%%%%%%%%%%%%%%%%%%%%%%%%%%%%%%%%%%%%%%%%%%%%%%%%%%%%%%%%

%%%%%%%%%%%%%%%%%%%%%%%%%%%%%%%%%%%%%%%%%%%%%%%%%%%%%%%%%%%%%%%%%%%%%%%%%%%
%%%%%%%%%%%%%%%%%%%%%%%%%%%%%%%%%%%%%%%%%%%%%%%%%%%%%%%%%%%%%%%%%%%%%%%

%%%%%%%%%%%%%%%%%%%%%%%%%%%%%%%%%%%%%%%%%%%%%%%%%%%%%%%%
%%%%%%%%%%%%%%%%%%%%%%%%%%%%%%%%%%%%%%%%%%%%%%%%%%%%%%%%
\section{Chern-Simon fluxes and their adiabatic addition}

The CF picture \cite{JainPRL89,JainPRB90} results from the
introduction of C-S fluxes. The vector potential $\vec a (\vec r)$
that produces  a flux of $\Phi=\alpha \Phi_0=\alpha hc/e$ is
\begin{equation}
\label{eq:CS_vect_pot} \vec a(\vec r)=\Phi\int
{d^2r^{\prime}\frac{\vec z \times (\vec r - \vec r^{\prime})} {|\vec
r - \vec r^{\prime}|^2}\rho(\vec r^{\prime})}
\end{equation}
where $\rho(\vec r)=\Psi^{\dagger}(\vec r) \Psi (\vec r)$ is the
density operator. The magnetic field resulting from the CS vector
potential is given by $\vec b (\vec r)=\Phi \sum_i\delta (\vec r
-\vec r_i)\hat z$ where $\vec r_i$ is the position of the
$i^{\rm{th}}$ electron.

A charged particle is moving on a circular path in $x-y$ having the
wavefunction $\Psi(\vec r)=e^{im\phi}u_m(r)$. When the flux tube is
added via a gauge transformation,  the eigenfunction is changed to
\begin{equation}
\overline{\Psi}(\vec r)=e^{-\frac{iq}{\hbar c}\int \vec a (\vec
r)\cdot d\vec r}\Psi(\vec r)~.
\end{equation}
Considering $\vec a(\vec r)= \hat \phi\Phi/(2\pi r) $, the new
wavefunction is $\overline{\Psi}(\vec r)=\exp (i(m-\alpha) \phi)
u_m(r)$. The CS flux does not influence the radial wavefunction, but
it introduces a phase of $-i\alpha \phi$. For relative motion of the
pair of charges, an odd integer $\alpha$  generates the famous
change of statistics from bosons to fermions and viceversa.
Introducing the CS flux, makes the many-body Schr\"{o}dinger
equation extremely complicated. A simplification occurs when the
mean-field approximation is made by replacing the density operator
by its ground state average density $n=<\Psi^{\dagger}\Psi>$ in the
expression of the CS vector potential. The particles move in an
effective magnetic field $B^*=B+\alpha\phi_0 n$ and the mean-filed
Coulomb interaction vanishes. The mean-field approximation
introduces an effective cyclotron energy scale $\hbar
\omega_c^{\ast}=(\hbar qB^{\ast})/(\mu c)$, which is irrelevant
since $q^2/\lambda_r<<\hbar \omega_c^{\ast}$.

Instead of making a gauge transformation we can introduce the CS
fluxes adiabatically \cite{QuinnQuinnPRB03}. We start with initial
electron pair state and slowly increase the value of CS flux. Such
technique leaves the phase of the wavefunction unchanged, but the
radial wavefunction is modified $u_m(r) \rightarrow u_{m+\alpha}(r)$
(in the absence of Coulomb interaction).

In the case of the system made of two particles with different
charges, the fluxes attached to the two particles are increased from
zero to $\alpha \phi_0$ and $\beta \phi_0$ respectively. The
Hamiltonian is
\begin{widetext}
\begin{equation}
\label{eq:genH} \hat H=\frac{1}{2m_1}\left\{\vec p_1-\frac{q_1}{c}
\left[ \vec A (\vec r_1) + \alpha \vec a(\vec r_1-\vec r_2)
\right]\right\}^2+\frac{1}{2m_2}\left\{\vec p_2-\frac{q_2}{c} \left[
\vec A (\vec r_2) -\beta \vec a(\vec r_1 - \vec r_2) \right]
\right\}^2+\frac{q_1 q_2}{|\vec r_1-\vec r_2|}~.
\end{equation}
where $a(\vec r)=\hat\phi(\phi_0/2\pi r) $. We suppose again that
the particles have the same specific charges. The Hamiltonian can be
decomposed into a CM part, an R part, and interaction between the CM
and R motions.
\begin{eqnarray}
\label{eq:HCM_CS} H_{CM}&=&\frac{1}{2M}\left[\vec P
-\frac{Q}{c}\vec A(\vec R)\right]^2+ \frac{\tilde q^2 \vec A^2(\vec R)}{2\mu c^2}~, \\
\label{eq:Hrel_CS} H_R&=&\frac{1}{2\mu}\left\{\vec p - \frac{q}{c}
\left[ \vec A (\vec r) +(\alpha+\beta)\vec a(\vec r)\right]\right\}
^2+ \frac{q_1q_2}{r} +
\frac{q_{\alpha}^2 \vec a^2(\vec r)} {2Mc^2}~,\\
\label{eq:Hint_CS} H_I&=& - \frac{q_{\alpha}}{Mc} \vec P \vec a(\vec
r)+\frac{q_{\alpha}q \vec A (\vec R)\vec a (\vec r)}{\mu c^2} ~.
\end{eqnarray}
where $q_{\alpha}=q_1 \alpha - q_2 \beta$
\end{widetext}
The CM and relative motions decouple completely if $q_{\alpha}=0$,
i.e. $q_1 \alpha =q_2 \beta$. This condition can be used to develop
the generalized CF picture for a plasma containing particles with
different charges. The relative motion Hamiltonian is very similar
to Eq. \ref{eq:Hrel}. The effective magnetic field is modified by
the CS flux. The new wavefunction will not change its phase, but the
orbital part is modified.

%%%%%%%%%%%%%%%%%%%%%%%%%%%%%%%%%%%%%%%%%%%%%%%%%%%%%%%%%%%%%%
%%%%%%%%%%%%%%%%%%%%%%%%%%%%%%%%%%%%%%%%%%%%%%%%%%%%%%%%%%%%%%%

%%%%%%%%%%%%%%%%%%%%%%%%%%%%%%%%%%%%%%%%%%%%%%%%%%%%%%%%%%%%%%%%
%%%%%%%%%%%%%%%%%%%%%%%%%%%%%%%%%%%%%%%%%%%%%%%%%%%%%%%%%%%%%%%%
\section{Generalized composite Fermion picture and the low-lying excitations
of the $\nu=5/2$ fractional quantum Hall state}

We introduce a generalized CF picture for a multicomponent plasma.
We think of every species of charged particle present in the plasma
as a having different color (red, blue, etc.). We attach to each
particle a flux tube carrying an integral number of different flux
quanta of different colors. Each charge will see only the flux tubes
having the same color, and no particles will see its own flux. For
example a ``red" charge will see the red flux tubes attached to all
other particles. The correlations between the particles having the
same color is identical to the one introduced in the original CF
picture. The generalized CF model also introduces  the correlations
between different type of particles. We need to obtain the same
correlations when ``blue" charges interacts with ``blue" fluxes on
``red" charges as when ``red" charges interact with ``red" fluxes
attached to blue charges. Adding $p q_{red}/e$ ``blue" fluxes to the
a ``red" charge causes the ``blue" charge to have exactly the same
blue-red correlations as adding $p q_{blue}/e$ ``red" fluxes to blue
charge. The CS charge times the CS flux must be the same.

In this section, we use the Haldane's spherical geometry
\cite{HaldanePRL83}, which maps the infinite planar surface onto a
spherical one, magnetic field being produced by a monopole placed in
the center of the sphere. The monopole strength is $2Q\Phi_0$ (where
$2Q$ is an integer) gives rise to a magnetic field $B=2Q\Phi_0/(4\pi
R^2)$ which is perpendicular to the spherical surface. The single
particle eigenstates are called monopole harmonics and denoted by
$|Q,\ell,m>$\cite{MonoploeHarmonics1,MonopoleHarmonics2}. They are
eigenfunctions of the square of the angular momentum operator
$\hat\ell^2$ and its $z-$projection $\hat\ell_z$ with eigenvalues
$\ell(\ell+1)$ and $m$ respectively, and $|m|\leq \ell$. Landau
levels are replaced by angular momentum shells $\ell=Q+n$, where $n$
plays the role of the LL index. The energy of the state $|Q,\ell,m>$
is $E_{\ell}=(\hbar \omega_C/2Q) [\ell(\ell+1)-Q^2]$. In order to
obtain the energy spectrum of an $N-$particle system, the Coulomb
interaction $e^2/r_{ij}$, is diagonalized in the non-interacting
basis set.

In this geometry, the effective monopole strength seen by CFs of
``color" $a$ will be
\begin{equation}
\label{eq:genQeff}
2Q_a^{\ast}=2Q-\sum_b(m_{ab}-\delta_{ab})(N_b-\delta_{ab})~.
\end{equation}
with $q_a m_{ab}=q_b m_{ba}$.

This model is applied to understand the lowest bands of states in
the case of $\nu=5/2$ state. It is clear that the correlations and
the elementary excitations are better understood for LL0 than for
LL1 and higher Landau levels. The Coulomb interaction is described
using the pseudopotentials $V(\mathcal R)$, where $\mathcal R$ is
the relative angular momentum $\mathcal R= 2\ell -L^{\prime}$,
$L^{\prime}$ being the pair angular momentum. It is well-known that
Laughlin correlations (the avoidance of pair states with small
values of $\mathcal R$) occur only when the pseudopotential
$V_n(\mathcal R)$ describing the interaction energy of a electron
pair with angular momentum $L'=2 \ell -\mathcal R$ in LLn is
``superharmonic", i.e. rises with increasing $L'$ faster than
$L'(L'+1)$ as the avoided value of $L'$ is approached
\cite{QuinnWoysSSC98,WojsQuinnSSC99, QuinnWPhysE00}. A
pseudopotential that is not ``superharmonic"  does not induce
Laughlin correlations \cite{SitkoYiQuinnPRB97} and instead results
in formation of pairs. In LL0 the pseudopotential is superharmonic
for all values of $\mathcal R$. The CF picture applied for electrons
in LL0, introduces $\ell^{\ast}= |\ell-p(N-1)|$, where $p$ is an
integer and explains that the lowest band of states will contain the
minimum number of QP excitations required by the values of $N$ and
$2\ell$ \cite{ChenQuinnPRB93} The QHs reside in the angular momentum
shell $\ell_{\rm{QH}}=\ell^{\ast}$; the QEs are in the shell
$\ell_{\rm{QE}}=\ell_{\rm{QH}}+1$.

\begin{figure}
\centerline{\includegraphics[width= 0.5 \linewidth]{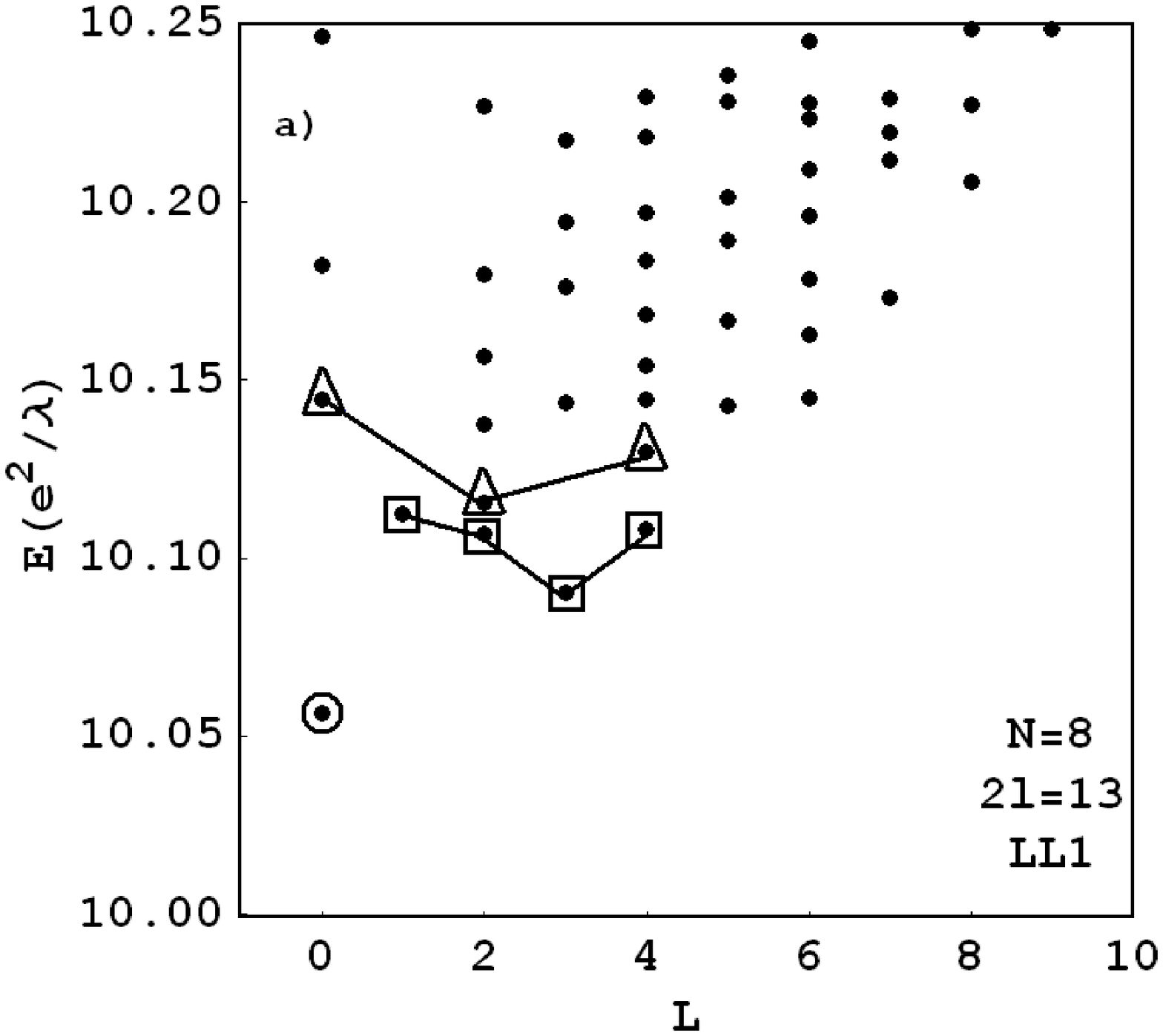}
\includegraphics[width= 0.5 \linewidth]{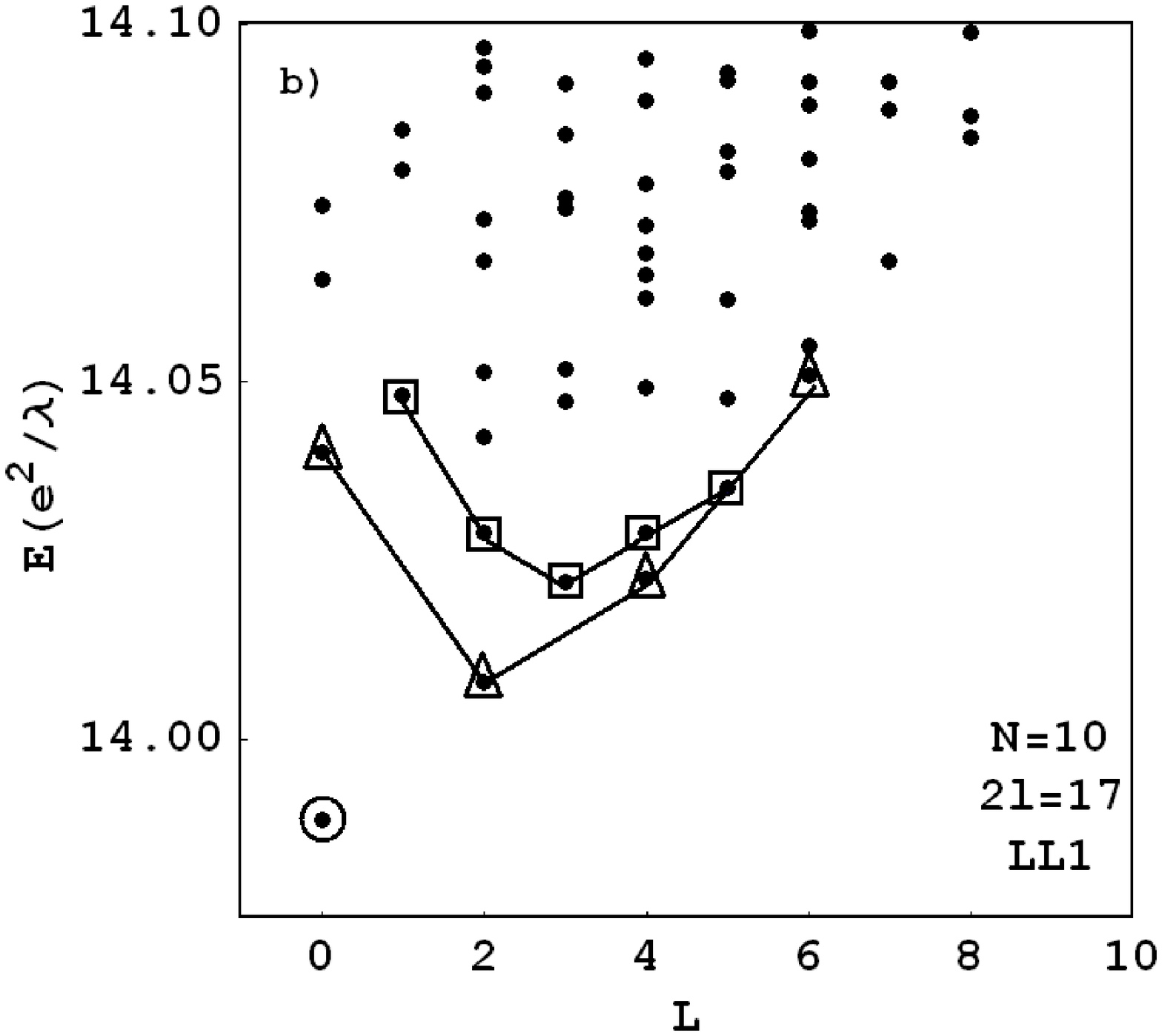}}
\centerline{\includegraphics[width= 0.5 \linewidth]{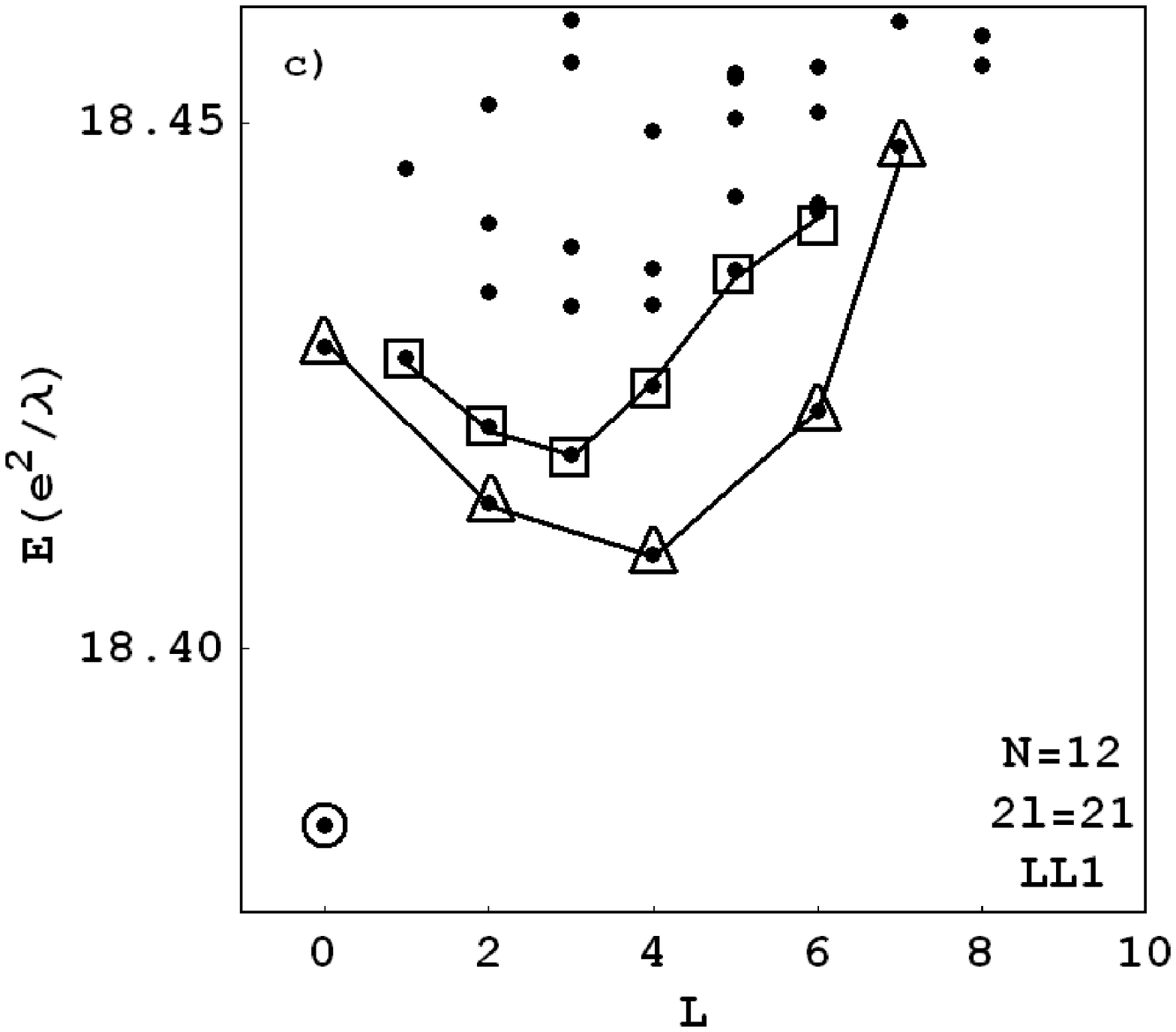}
\includegraphics[width= 0.5 \linewidth]{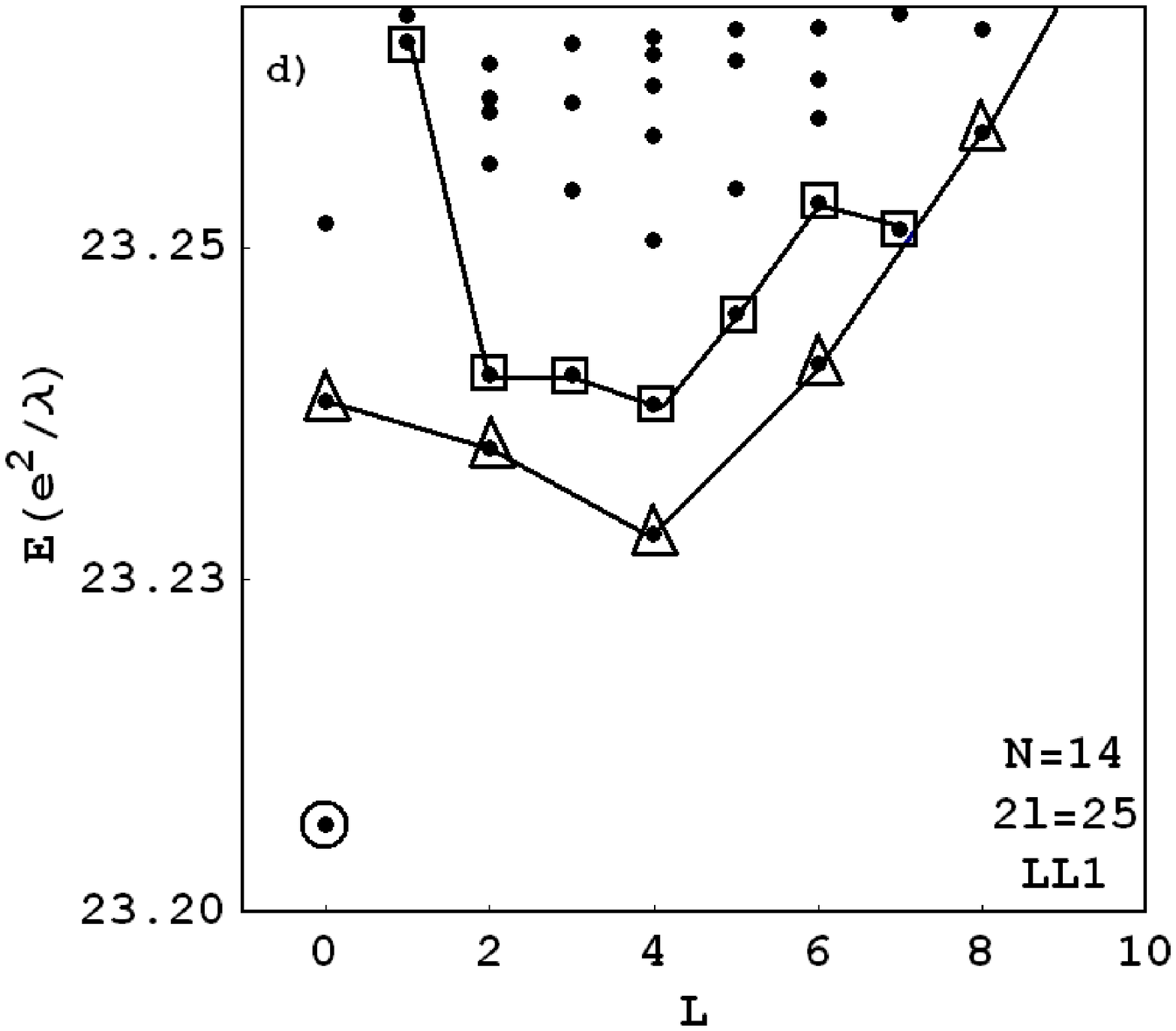}}
\caption{The energy spectra of electrons in LL1 for $2\ell=2N-3$
generating incompressible ground states. Circles represent the IQL
states, triangles represent the pair magnetoroton states, and
squares represent the ``broken-pair" states.} \label{Fig:nu52states}
\end{figure}

In LL1, the pseudopotential is only ``weakly" superharmonic
\cite{WojsPRB01, WojsQuinnPRB05} for $R=1$ and as a consequence it
does not support Laughlin correlations at $1/3\leq \nu_1 \leq
2/3$\cite{prlarxiv}. The correlations can be described in terms of
the formation of $N_{\rm{P}}=N/2$ pairs\cite{prlarxiv} when $N$ is
even. The electrons tend to form pairs with $\ell_{\rm{P}}=2\ell-1$.
To avoid violating the exclusion principle, we can't allow Fermion
pairs (FPs) to be too close to one another. We do this by
restricting the angular momentum of two pairs to values less than or
equal to
\cite{QuinnWojsYiPLA03,WojsWodzinskiQuinnPRB05,WojsWodzinskiQuinnPRB06,QuinnQuinnSSC06}
\begin{equation}
\label{eq_lFP} 2\ell_{\rm{FP}}=2\ell_{\rm{P}}-3(N_{\rm{P}}-1)~,
\end{equation}
implying that the FP filling factor satisfies the relation
$\nu_{\rm{FP}}^{-1}=4\nu_{1}^{-1}-3$. The factor of 4 is a
reflection of $N_{\rm{P}}$ being half of $N$ and the LL degeneracy
$g_{\rm{P}}$ of the pairs being twice $g$ for electrons.
Correlations are introduced through a standard CF transformation
applied to FPs:
\begin{equation}
\label{eq:lFPeeff} 2\ell_{\rm{FP}}^*=
2\ell_{\rm{FP}}-2p_{\rm{P}}(N_{\rm{P}}-1) ~,
\end{equation}
Selecting $2p_{\rm{P}}=4$ results in
$2\ell_{\rm{FP}}^*+1=N_{\rm{P}}$ and pairs forming a $L=0$
incompressible quantum liquid (IQL) ground state. This occurs if the
number of electrons and the angular momentum shell satisfy the
relation $2\ell=2N-3$ (or the electron-hole conjugate $2\ell=2N+1$).
These IQL states are marked by circles in Fig. \ref{Fig:nu52states}.

In order to understand the lowest bands of states we will assume two
types of elementary excitations. The first type will consists of an
empty FP state in the lowest FP level [a quasi-hole FP (QHFP) with
angular momentum $\ell_{\rm{FP}}^{\ast}$] plus one filled FP state
in the first excited FP level [a quasiparticle FP (QPFP) with
angular momentum $\ell_{\rm{FP}}^{\ast}+1$]. Since
$2\ell_{\rm{FP}}^{\ast}=(N_{\rm{P}}-1)$, this process give rise to a
``magnetoroton" state of a QHFP with
$\ell_{\rm{QHFP}}=(N_{\rm{P}}-1)/2$ and a QPFP with
$\ell_{\rm{QPFP}}=(N_{\rm{P}}+1)/2$. The resulting``magnetoroton"
band has $L=1\oplus 2 \oplus \ldots \oplus N_{\rm{P}}$. This band is
marked by squares in Fig. \ref{Fig:nu52states}. Up to an overall
constants this band represents the interaction pseudopotential of a
QHFP and a QPFP as a function of the total angular momentum.
Unfortunately the width of the band is not small compared to the
minimum gap required to produce a magnetoroton of FPs, so it is not
as useful as the QE and QH pseudopotential obtained from Laughlin
correlated states in LL0 which contain a pair of QEs or a pair of
QHs \cite{ChenQuinnPRB93}.

%%%%%%%%%%%%%%%%%%%%%%%%%%%%%%%%%%%%%%%%%%%%%%%%%%%%%%%%%%%%%%%%%%%%%%%%%%%
%%%%%%%%%%%%%%%%%%%%%%%%%%%%%%%%%%%%%%%%%%%%%%%%%%%%%%%%%%%%%%%%%%%%%%%

Another possible band of low lying excitations could result from
breaking one of the CF pairs into two constituent unpaired
electrons, each with charge $-e$ and angular momentum $\ell$. We
propose to treat the system of $N_{\rm{P}}^{\prime}=N_{\rm{P}}-1$
FPs and the $N_e=2$ unpaired electrons by our generalized CF
picture. Let the $2\ell_{\rm{FP}}^{\prime}=
2\ell_{\rm{P}}-3(N_{\rm{P}}^{\prime}-1)$. Then the following
equations describe correlations of the $N_{\rm{P}}^{\prime}$ FPs and
$N_e=2$ unpaired electrons:
\begin{equation}
2\ell_{\rm{FP}}^{\prime \ast }=2\ell_{\rm{FP}}^{\prime}-2p_{\rm{P}}
(N_{\rm{P}}^{\prime}-1)- 2\gamma N_e~. \label{eq:FP_FP_e_corr}
\end{equation}
\begin{equation}
2\ell_{e}^{\ast}=2\ell-2p_e (N_e-1) - \gamma N_{\rm{P}}^{\prime}~.
\label{eq:e_FP_e_corr}
\end{equation}

Eq. \ref{eq:FP_FP_e_corr} tells us that the effective angular
momentum of one FP is decreased from $\ell_{\rm{FP}}$ by
$p_{\rm{P}}$ times the  number of other FPs and by $\gamma$ times
the number of unpaired electrons. Eq. \ref{eq:e_FP_e_corr} tells us
that the effective angular momentum of one unpaired electron is
decreased by $p_e$ times the number of other unpaired electrons and
by $\gamma /2$ times the number of CF pairs. Note that
$2p_{\rm{FP}}$ and $2p_e$ are even, and that $\gamma$ can be odd or
even.  Equations \ref{eq:FP_FP_e_corr} and \ref{eq:e_FP_e_corr}
define the generalized CF picture in which different types of
Fermions, distinguishable from one another, experience correlations
which leave them as Fermions (since $2p$ is even) and give the same
correlations between members of two different species since the
product of CS charge and the CS flux added are the same (i.e $-e
\cdot 2\gamma = -2 e \cdot \gamma$).

When using $p_{\rm{P}}=2$, $p_e=1$,  $\gamma=2$,
$N_{\rm{P}}^{\prime}=N_{\rm{P}}-1$ and $N_e=2$, and
$2\ell_{\rm{FP}}^{\prime}=2\ell_{\rm{P}}-3(N_{\rm{P}}^{\prime}-1)$,
eq. \ref{eq:FP_FP_e_corr} give the FPs effective angular momentum as
$\ell_{\rm{FP}}^{ \prime \ast} =N_{\rm{P}}-2$ and the FPs will form
an IQL $L=0$ state. The 2 electrons become CFs occupying the level
$2\ell_{e}^{\ast}=N-3$. They generate the ``broken pair" band of
states at $L=0 \oplus 2 \oplus \cdots \oplus N-4$. This band of
states is represented by triangles in Fig. \ref{Fig:nu52states}.

Some of these states were discussed by Greiter et al.
\cite{GreiterWenWilczekPRL91, GreiterWenWilczekNuclPhys92} but not
in terms of a generalized CF picture capable of predicting the
allowed values of $L$ in the lowest band of energy levels.

%%%%%%%%%%%%%%%%%%%%%%%%%%%%%%%%%%%%%%%%%%%%%%%%%%%%%%%%%%%%%%%%%%5
%%%%%%%%%%%%%%%%%%%%%%%%%%%%%%%%%%%%%%%%%%%%%%%%%%%%%%%%%%%%%%%%%%5

\section{Conclusions}

We developed a generalized CF model in which a plasma contains
particles with different charges. The adiabatic introduction of CS
fluxes was used. It is worth noting that for the generalized CF
picture the correlations between a pair of particles can be thought
of as resulting from adiabatic addition of fictitious CS flux quanta
to one particle that is sensed by fictitious charge on the other. We
applied our model to the system with $\nu=5/2$ filling factor. Our
interpretation is an attempt to understand some of the low lying
excitations of the $\nu_1=1/2$ state in a simple CF type picture.
The numerical data confirm our model. Our results do not contain
degenerate QP states that give rise to non-Abelian statistics.
Further studies with realistic electron pseudopotentials should be
undertaken to see if non-Abelian quasiparticles appearing at values
of $2\ell$ slightly different than $2N-3$  are only an artifact of
the Greiter, Wen and Wilczek model interaction.

%%%%%%%%%%%%%%%%%%%%%%%%%%%%%%%%%%%%%%%%%%%%%%%%%%%%%%
%%%%%%%%%%%%%%%%%%%%%%%%%%%%%%%%%%%%%%%%%%%%%%%%%%%%%%

\bibliography{RMPQuinn}
\end{document}